\documentclass[fleqn,twoside]{article}
\usepackage{espcrc2}

\usepackage{epsfig}
\usepackage[figuresright]{rotating}

\newcommand{\AmS}{{\protect\the\textfont2
  A\kern-.1667em\lower.5ex\hbox{M}\kern-.125emS}}

\title{Alternative to Domain Wall Fermions}

\author{H. Neuberger\address[RUPHY]{Department of Physics and Astronomy, \\
        Rutgers University,\\ Piscataway, NJ 08855, USA}\thanks{Permanent Address}
        \address{Univ. Paris VI, \\ T.16,  4 Place Jussieu, $1^{\rm er}$ etage,\\
        75252 Paris Cedex 05, France}}

\begin{document}

\begin{abstract}
An alternative to commonly used domain wall fermions is presented.
Some rigorous bounds on the condition number of the associated linear problem
are derived. On the basis of these bounds and some experimentation 
it is argued that domain wall fermions will in general be associated with
a condition number that is of the same order of magnitude as
the {\it product} of the condition number of the
linear problem in the physical dimensions by the inverse bare quark mass.
Thus, the computational cost of implementing true domain wall
fermions using a single conjugate gradient algorithm 
is of the same order of magnitude as that of implementing the overlap 
Dirac operator directly using two nested conjugate gradient algorithms. 
At a cost of about a factor of two 
in operation count it is possible to make the memory
usage of direct implementations of the overlap Dirac operator independent
of the accuracy of the approximation to the sign function and of the same
order as that of standard Wilson fermions. 

\vspace{1pc}
\end{abstract}

\maketitle

\section{INTRODUCTION}

This review will start with a sketch of the kinematical-algebraic aspects
of the overlap Dirac operator in the vector-like context. Next comes a general
discussion of numerical implementations of the overlap Dirac operator. 
Section 2 is devoted
to an alternative domain wall model. This model is domain-wall like in the
sense that an extra dimension is added and the computation of the light fermion
propagator requires a single conjugate gradient procedure, albeit for a matrix
representing fermions in five dimensions. On the other hand, the model is designed
so that its output in exact arithmetic is the same as that of an iterative method 
implementing the overlap Dirac operator directly.  The latter method requires
a two-level nested conjugate gradient procedure. In section 3 rigorous results
on spectral properties of 
our model are presented and in section 4 these results are compared to true
domain wall fermions. The main conclusion is that nested procedures typically are more
efficient than implementations based on 
domain-walls. This counter-intuitive conclusion is explained
by the condition number of the domain wall problem being the product of the
condition number of the four dimensional problem by the inverse bare quark mass. 
The latter two factors govern individually the two nested cycles in direct 
implementations. 
In Section 5 it is shown that one can eliminate the requirement of linearly growing
memory consumption for increasing accuracy at the cost of a factor of two in operations.
In practice, the factor of two is often not felt because the implementation is memory bound.
Section 6 contains some final comments. 

A large part of this talk is about work done in collaboration with
Rajamani Narayanan~\cite{alt}.

\subsection{Algebraic structure}

The overlap formulation~\cite{overlap} of vector-like gauge theories on the lattice
preserves chiral symmetries exactly, a property thought to be unattainable
for many years. If one adds to the Ginsparg Wilson relation, as originally
formulated in 1982~\cite{gw}, a requirement
of $\gamma_5$-hermiticity, the combination is equivalent to the overlap at the
algebraic level. The set-up has been of interest to mathematicians much earlier
as a generalization of the concept of angle between two straight intersecting
lines in a plane. The plane is generalized to a vector space over the complex numbers
and the two lines to two subspaces of the vector space~\cite{kato}. 
In our application the dimension of the vector
space will always be even, but the subspaces can have unequal dimensions - when
they do, the angle concept looses its meaning. 
The Kato setup is also meaningful
for infinite dimensional Hilbert spaces and has other 
applications to physics~\cite{avron}. 

The subspaces are defined by projectors onto them, in our case the projectors are
hermitian and replaced by linearly related reflections, $\epsilon$ and 
$\epsilon^\prime$. $\epsilon^2=1$ and $\epsilon^{\prime~2}=1$. The $-1$ eigenspaces
of $\epsilon$ and $\epsilon^\prime$ are the subspaces in question. They are
spanned by orthonormal sets denoted by $\{ v_k \}$ and $\{v^\prime_{k^\prime} \}$
with $\epsilon v_k + v_k =\epsilon^\prime v^\prime_{k^\prime} + v^\prime_{k^\prime} =0$.
The information about the 
relative positioning of the subspaces is contained in the overlap matrix
$M$:
\begin{equation}
M_{k^\prime k} = {v^\prime_{k^\prime}}^\dagger v_k
\end{equation}
The coarsest measure of relative orientation is obviously $|\det M|$.
The main identity~\cite{overlap} is
\begin{equation}
|\det M |^2=\det D_o,~~~~D_o=\frac{1+\epsilon^\prime \epsilon}{2}
\end{equation}
In our case $\epsilon^\prime=\gamma_5$ and $\epsilon={\rm sign} (H_W)$, where
$H_W=\gamma_5 D_W$ and $D_W$ is any lattice version of the Dirac operator
describing fermions with negative mass of order $\frac{1}{a}$ where $a$ is
the lattice spacing. $\epsilon^\prime$ can be thought of as describing Dirac fermions of 
positive infinite mass. $D_W$ is restricted by the requirement that $H_W$ be
hermitian. The simplest choice, which minimizes operations in the computation
of the action of $D_W$ on a vector, is to pick $D_W$ as the Wilson lattice 
Dirac operator. It is possible that when all practical aspects of a simulation
are taken into account a seemingly more complicated choice would pay better off.

$H_W$ depends on the background gauge field represented by a collection of
unitary link matrices $U_\mu$. $H_W$ has the following properties:
\begin{itemize}
\item $H_W$ is too large to be stored in the computer 
memory in its entirety, but
it is sparse, so its action on vectors can be implemented.
\item The spectrum of $H_W$ is bounded by a finite bound 
that does not depend on the gauge
background.
\item If all products of gauge matrices on links around plaquettes, $U_p$, obey
$||U_p-1||<\eta$, where $\eta$ is a small positive number independent of the gauge
background, the spectrum of $H_W^{-1}$ is also bounded by a finite number, 
independently of the gauge-field. 
\end{itemize}

\subsection{Basic numerical issues}

The calculation of the action of $\epsilon$ on a vector must
use sparse matrix techniques. The boundedness of the spectrum
means that the sign function needs to be approximated accurately
only in two finite segments symmetrical about zero, which contain
the entire spectrum of $H_W$, for any gauge background allowed by
the pure gauge action. The main strain on the approximation occurs
around zero, where the gap turns out in practice to be very small.
This is where essentially all the numerical cost goes. 

There are two main approaches to the approximate implementation of
the sign function: One is the direct approach (overlap), and the
other is indirect (domain wall fermions). 

In the direct approach one
looks for a rational approximation for the sign  function in the
range defined by the bounds on the spectrum of $H_W$. The rational
approximation is written as a sum of pole terms. The crucial point
is that the action of each pole term on a vector need not be calculated 
separately: rather, the action of all terms can be calculated 
simultaneously, in one single pass through the conjugate gradient 
algorithm~\cite{shifttrick}. The approximation gets more accurate when the number of
pole terms is increased. Settling for a certain number of terms, $n$,
one obtains an approximation for 
$\epsilon$, 
\begin{equation}
\epsilon^n (H_W) =\sum_{s=1}^n \frac{w_s}{H_W^2 + m_s^2} 
\end{equation}
The ``masses'' $m_s^2$ have to be non-negative, but the weights $w_s$
can take either sign. In practice we need the action of $\frac{1}{D_o}$
on a vector, so the action of $\epsilon^n$ is needed many times for one
evaluation. We end up with two levels of nested conjugate gradient algorithms. 

In the context of domain wall fermions an approximation for the sign function 
is not constructed directly. Rather, one invents a larger problem, defined
by a matrix $H$. $H$ is determined by $H_W$ but its dimensions
are $n$-times larger. $H$ is still sparse. One then arranges that
in a given subspace of the larger space $H$ acts on one has
\begin{equation}
\left ( \frac{1}{H}\right )_{\rm specific~subspace} \approx \frac{1}{H_o},~~~H_o=\gamma_5 D_o
\end{equation}
Above, $D_o$ and $H_o$ can differ from eq. (2) by terms that disappear in the
continuum limit. The exact form is a matter of convenience. 
One ends up needing to do only one inversion,
using a single conjugate gradient algorithm (CG). In the most common applications
of domain wall fermions one uses a construction of $H$ 
due to Kaplan~\cite{overlap}. Historically,
Kaplan's formulation came first, and was a prime motivator for subsequent developments. 

\subsection{Motivation for model}

The main difference between the two methods is that one has a nested CG
in one and a single CG in the other, but employing a larger matrix. The objective
of my work with Narayanan~\cite{alt} was to invent a model containing a version of $H$ which
is close to Kaplan's domain wall fermions, but also to the rational approximation
so that a comparison of numerical costs may be carried out using rigorous methods. 
We wanted the output of either method to be the same in exact arithmetic, so
it would be only the ways of doing the calculation that differ. One way is similar to 
domain wall fermions and the other is direct.  

\section{THE ALTERNATIVE MODEL}

Let $\psi$ be the Dirac
field describing a light quark. We wish to end up with an effective action for $\psi$
given by:
\begin{equation}
S_{\rm eff} (\psi) = - \bar\psi \left [
{{1+\mu}\over 2} \gamma_5 + 
{{1-\mu}\over 2}\epsilon^n (H_W ) \right ]\psi .
\end{equation}
To be specific, we choose
\begin{equation}
\epsilon^n (H_W ) = {1\over n } \sum_{s=1}^n {1\over {c_s^2 H_W +
{s_s^2 \over H_W}}}
\end{equation}
with
\begin{eqnarray}
c_s= \cos \theta_s ,~~s_s=\sin \theta_s,~~~\theta_s =
{\pi\over{2n}} (s-{1\over 2})\nonumber\\
s=1,2,...,n .
\end{eqnarray}
This rational approximation can be easily replaced by others.

We now add $2n$ extra fields, $\bar\chi_s, \chi_s$ and $\bar\phi_s, \phi_s$,
pack them together with the light field into a combined field,  
$\bar\Psi = (\bar\psi, \bar\chi_1 ,
\bar\phi_1,...,\bar\chi_n,\bar\phi_n )$. The total action is
\begin{equation}
S= \bar \Psi H \Psi
\end{equation}
Introduce $a_n=\sqrt{{1-\mu}\over {2n}}$
and $b=-{{1+\mu}\over 2}$. The matrix $H$ is given by:
\begin{eqnarray}
\pmatrix{ 
b \gamma_5 & a_n &0
&\dots
&a_n&0\cr
a_n& c_1^2 H_W & s_1 &\dots &0&0\cr
0& s_1 &-H_W  &\dots &0&0\cr
a_n&0&0&\dots& 0& 0\cr
0&0&0&\dots&0&0\cr
\vdots& \vdots& \vdots&\dots&0&0\cr
a_n &0&0&\dots&c_n^2 H_W & s_n\cr
0&0&0&\dots&s_n&-H_W\cr
}\nonumber\\\nonumber
\end{eqnarray}
Our goal is attained by the model because of the identity:

\begin{eqnarray}
\int d\bar\Psi d\Psi e^{\bar\Psi H \Psi} = 
\left [ \prod_{s=1}^n \det (c_s^2 H_W^2 +s_s^2 )
\right ] \times \nonumber\\
\int d\bar\psi d\psi e^{-\bar\psi \left ( {{1+\mu}\over 2}
\gamma_5 +{{1-\mu}\over {2n}}\sum_{s=1}^n
{1\over {c_s^2 H_W + {{s_s^2}\over H_W}}} \right ) \psi }
\end{eqnarray}
The prefactor can be canceled by adding pseudofermions, which will
be decoupled in the $s$ index. 

Let us now roughly estimate computational costs.
In the direct approach the number of inner iterations is 
approximately given by the condition number of $H_W$,
$\kappa (H_W )$ (which
is the square root of the condition number of $H_W^2$). 
The number of outer iteration goes as the condition number
of $D_o$ which is roughly $\frac{1}{\mu}$. Thus, the total
number of $H_W \psi$ actions is roughly given by 
$\frac{\kappa(H_W)}{\mu}$. 

In the domain wall version the number of $H\Psi$
operations is governed by $\kappa (H)$. Every $H\Psi$
operation counts roughly as $2n$ $H_W$ operations. We need
to estimate $\kappa (H)$ in terms of $\kappa (H_W)$ and $\mu$.
To find $\kappa (H)$ we need the maximal and minimal eigenvalues
of $H^2$. The basic trick for finding $\lambda_{\rm max} (H^2)$
and $\lambda_{\rm min} (H^2)$ is the derivation
of an exact formula for $\det (H-z)$.
\begin{eqnarray}
\det (H-z) = \nonumber\\
\prod_{s=1}^n 
\det \left [ (c_s^2 H_W -z )(H_W +z ) +s_s^2 \right ] \times
\nonumber\\
\det \left [ {{1+\mu}\over 2}\gamma_5 +z 
+{{1-\mu}\over 2} f_n (H_W ,z )\right ]
\end{eqnarray}
Here,
\begin{equation}
f_n (H_W , z ) = {1\over n} \sum_{s=1}^n {1\over {c_s^2 H_W +
{s_s^2 \over {H_W +z }} -z}} .
\end{equation}
Eigenvalues of $H$ are roots of the equation $\det (H-z)=0$.
All the roots come from roots of the last factor. (Roots of the
factors in the product over $s$ are canceled by poles in the last
factor. So, the spectrum of $H$ is determined by the last factor.) 
We write
\begin{equation}
f_n (H_W , z ) = S_n(H_W -z , {1\over{H_W +z }} -z ).
\end{equation}
where,
\begin{equation}
S_n(a,b) = {1\over n} \sum_{s=1}^n {1\over {a c_s^2 +b s_s^2}},
\end{equation}
The point is that one has explicit formulae for $S$, so long
$a$ and $b$ are real. 
If $ab >0$ we have
\begin{eqnarray}
S_n(a,b) = {{{\rm sign} (b)}\over \sqrt{ab}} \tanh (n\omega)
\nonumber\\
\omega = \log\left ( |1+\sqrt {a/b} | /|1-\sqrt {a/b }| \right ) .
\end{eqnarray}
If $ab <0$,
\begin{eqnarray}
S_n(a,b) = {{{\rm sign} (b)}\over \sqrt{-ab}} \tan (n\omega),~~~
0\le \omega \le {\pi\over 2},\nonumber\\ 
e^{i\omega} = ( 1+i\sqrt {-a/b} )/(1-i\sqrt{-a/b}) .
\end{eqnarray}
These formulae make the $n$-dependence explicit.

\section{RIGOROUS RESULTS}

Let me describe the
main idea for deriving bounds on the eigenvalues of $H$.
We are looking for zero modes of:
\begin{eqnarray}
X(z)=\frac{1+\mu}{2}\gamma_5 +z +\frac{1-\mu}{2} 
f_n (H_W ,z )\nonumber\\
\equiv \frac{1+\mu}{2} \gamma_5 + X^\prime (z)
\end{eqnarray}
Conditions on $z$, depending on $\mu$ and spectral properties
of $H_W$, limit the eigenvalues of $X^\prime (z)$ to some
range. Suppose $\Psi_0$ is a zero mode of $X (z)$. It can exist
only if $\left | \frac{1+\mu}{2} \right |^2$ is in the range
of eigenvalue of $\left ( X^\prime (z)\right )^2$. But,
$f_n (H_W , z)$ is bounded from above and below for $z=\pm\infty$ 
and for $z=0$. So, it is possible to exclude 
vicinities of $z=0$ and $|z|=\infty$.

In this way we obtain rigorous bounds:
\begin{equation}
\lambda_{\rm max}^{\frac{1}{2}} (H^2 )\le \sqrt {||H_W||^2 +2 } 
\end{equation}
and
\begin{equation}
\lambda_{\rm min}^{\frac{1}{2}} (H^2) \ge 2\mu 
\lambda_{\rm min}^{\frac{1}{2}} (H_W^2) \rho_n (H_W, \mu)
\end{equation}
$\rho_n$ is a function of the spectrum of $H_W$ and of
$\mu$. It is well defined and calculable but clumsy to write
down. For ranges of practical interest $\rho_n \sim 1$. 

This leads us to:
\begin{equation}
\kappa (H) \le {{\kappa (H_W )} \over {2\mu}} 
{{\sqrt {1 +{2\over {||H_W||^2}}}} \over 
{\rho_n (H_W, \mu )}
}
\end{equation}
where the last factor is close to unity in practice.

The meaning of the result is quite obvious: We need to overcome both
the lattice artifact of having at times almost zero modes for $H_W$
and the physical effect of small quark mass ($\mu$, when close to zero,
gives the bare lattice quark mass). 

Clearly our conclusion speaks in favor of the direct approach, if
we assume the bounds to be typically saturated (the action
of $H$ is $2n$ more expensive than that of $H_W$). But, before
jumping to this conclusion we should try to decide how good
our bounds are. It turns out they are quite good.
The bound on the $\lambda_{\rm max} (H^2)$ is typically saturated. 
As far as $\lambda_{\rm min} (H^2 )$ goes we can prove
\begin{eqnarray}
\lambda_{\rm min} (H^2 ) \le \nonumber\\
\lambda_{\rm min} \left (
\left [   {{1+\mu}\over 2}\gamma_5 + 
{{1-\mu}\over 2} \varepsilon_n (H_W ) \right ]^2 \right )
\end{eqnarray}
One cannot do better than $\lambda_{\rm min} (H_W^2 )$ either.
For $\mu$ small enough and $n$ large enough
\begin{equation}
\lambda_{\rm min} (H^2 ) \le \lambda_{\rm min} (H_W^2 ) 
+\sin^2 \frac{\pi}{2n}
\end{equation}
Thus either a very small eigenvalue of $H^2_W$ or a very small mass
are guaranteed to make the smallest eigenvalue of $H^2$ very small too. 
The lower bound in eq. (18), if optimal, indicates that it is possible for the smallest
eigenvalue of $H^2$ to be as small as the product of these two small numbers.

\section{COMPARISON TO DOMAIN WALL FERMIONS}

For true domain wall fermions with action $D$ we can 
prove only one kind of bound:
\begin{eqnarray}
\lambda_{\rm min} (D^\dagger D) \le \left [ \lambda_{\rm min}^{1/2} (H_W^2 )
+{{1+\mu}\over n} \right ]^2 \nonumber\\
+(1+\mu)^2 \left ({1\over n} -{1\over {n^2}}
\right )
\end{eqnarray}
An approximate analysis yields 
\begin{equation}
\lambda_{\rm min}^{\frac{1}{2}} (D^\dagger D) 
\le 2\mu \lambda_{\rm min}^{\frac{1}{2}} (H_W^2)
\end{equation}
Upper bounds on $\lambda_{\rm max} (D^\dagger D)$ are of order 10
in practice. 

We conclude that the matrices describing the version of
domain wall fermions used in large scale simulations have conditions
numbers that behave similarly to the condition number of our 
model. 

\begin{figure}
%\centerline{{\setlength{\epsfxsize}{3in}\epsfbox[20 20 600 600]{hop.ps}}}
\epsfxsize=2.5in
\epsfbox[20 20 600 600]{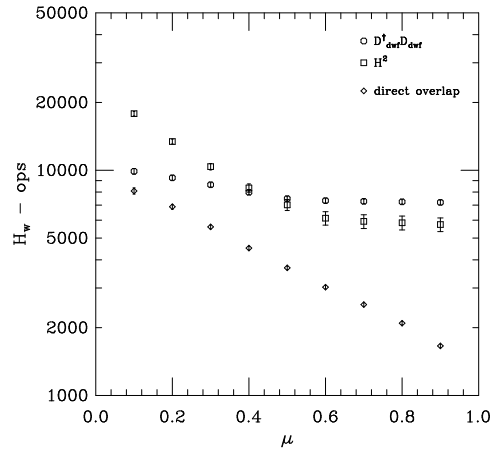}
\caption{ A comparison of the number of operations of $H_W$ 
for the inversion of the fermionic operator in three cases:
true domain wall fermions, the direct implementation of the rational approximated
sign function and 
our alternative domain wall fermions.
}
\end{figure}

In a modest numerical experiment on a two-dimensional $U(1)$ model
we compared domain wall fermions using our alternative model,
true domain wall fermions and the direct overlap approach. The
pure gauge action was of the single plaquette type, with a lattice
coupling $\beta=4$ and the lattice size was $8\times 8$. 

We performed
the calculations needed to obtain the 
condensate $\langle \bar \psi_{\rm phys}\psi_{\rm phys} \rangle$.
We required the norm of the residual to go down to $10^{-8}$ and counted
operations of $H_W$ on a vector. The value of $n$ used was $20$. We
did not use preconditioning in any of the three methods. Results were 
obtained using 20 gauge field configurations. Figure 1 shows the number of
$H_W$-on-vector operations as a function of quark mass. 

The comparison between the direct overlap and the alternative
method is simple because, by design, the numbers one would get in exact arithmetic
are the same in either method. Comparing to true domain wall fermions is more
difficult because one needs to match parameters to get similar physics, and this
is ambiguous. For example we set the Wilson mass parameter in $H_W$ to $-1.5$
in all three cases and used the same mass parameter $\mu$ although this certainly
is not correct for very small $\mu$.

We found that the best was to use the direct overlap approach, but this
experiment was very limited and one should not immediately
draw conclusions about numerical QCD. However, there is enough
evidence by now to scrutinize seriously the question
whether it is worth investing large scale computational resources
into true domain wall fermions given that there is a direct overlap
alternative. In particular we have employed no preconditioning
and ignored many other options of optimization. 

Theoretically, one is more comfortable with overlap
fermions, since they differ from other four dimensional fermions
just by a more complicated action. If the domain wall approach
were more effective computationally one could have the best of the
two worlds by using our alternative domain wall fermions. Whichever
the case may be, it is hard to argue in favor of true domain wall fermions.

\section{DOUBLE PASS ALGORITHM}

Until now we concerned ourselves only with counting operations.
But, as is well known, this is only one aspect of a computation.
Memory requirements and access patterns matter as much, or more, depending
on architecture. Until now all methods required a factor of $n$ times
more memory than on ordinary Wilson fermion code would. 

In the direct approach there is an option to trade a factor of order 2
in operations for reducing the memory needed to that of ordinary Wilson
fermions. Basically, this is possible because the heart of the
code is a conjugate gradient procedure, and only a certain linear
combination of the $n$ vectors that are acted on is needed. 

The conjugate
gradient procedure is closely related to the Lanczos scheme for bringing a hermitian
matrix to tri-diagonal form. It is well known that in the Lanczos scheme, if one
wishes additional information (for example one needs an eigenvector) one can
avoid large memory consumption by going through the algorithm twice,
once to collect the coefficients and a second time to accumulate the needed
data. The same works here~\cite{twopass}. 

In the multishift scheme~\cite{shifttrick} one iterates over an index $i$
updating $n$ vectors $x^s_i$, starting from an input vector $b$. 
At each iteration these 
vectors are determined by a few $s$-dependent scalars and by three $s$-independent
vectors that make up the core conjugate algorithm, $w,r,p$. Among them,
$r$ is the residual. When the iteration is stopped we have $n$ $x^s$ vectors
and they are summed into $y$, our approximation to the vector $\epsilon b$
\begin{equation}
y=\sum_{s=1}^{n} w^s x^s
\end{equation}
Actually, the iteration must be effectively stopped 
after a different number of steps for each $s$,
since for higher values of $s$ the ``mass'' is larger and, as a result, the convergence
much faster. Clearly, $y$ is made out of the basic Krylov vectors generated
in the core conjugate gradient; the single reason we need to store $n$ vectors
is that the components of $y$ in the Krylov basis at step $i$ in the iteration
are not yet known because they depend also on steps $i+1,i+2,\dots$. 

The idea is now to use a first pass to calculate the needed conjugate gradient 
scalars which are $i$ dependent but not $s$-dependent. 
Using these scalars
we can now compute in an iteration storing scalars only the extra $s$ 
and $i$-dependent scalars needed for implementing the multishift. 
It is possible now to also compute a set of $s$-independent scalars $R_i$
which have the property that
\begin{equation}
y_{i+1}=\sum_{k=0}^{i+1} R_k r_k
\end{equation}
becomes at the last iteration the desired approximation for $y$. But, we need
the vectors $r_i$ again so we need to run the basic conjugate gradient again.
Hence, a second pass is required, but one needs to store only four large vectors
for any $n$.

A more detailed
description of the algorithm can be found in~\cite{twopass}. 
The surprise was that with a code written in a higher level language
the two pass version can actually run faster than the single pass version,
by about 30 percent. In the two pass version both the operations count
and the memory usage are independent on $n$ ! The speed up is certainly a surprise
and must be strongly machine dependent. But, it turns out, that one gets
the same amount of speed-up on an SGI O2000~\cite{twopass}
as on a Pentium III PC~\cite{Narkis}.

\section{FINAL COMMENTS}

Let us summarize roughly the situation we were looking at when we began
to work on reference~\cite{alt}: Although an approach based 
on the overlap Dirac operator
looked theoretically cleaner, true domain wall fermions were more attractive
numerically. Our analysis has led us to the conclusion that there
is no evidence that true domain wall fermions have even a numerical advantage. 

In all cases we looked at, one faces a problem related to almost zero modes
of $H_W$. This requires large numbers of extra fields in order
to preserve chirality. It also
affects adversely the condition numbers. Whichever method we
use, the worst case condition numbers are a product of the inverses
of two main scale ratios: The first is the scale of the small eigenvalues
of $H_W^2$ divided by an upper bound of the order of 5-10 in lattice
units. The second scale ratio is the lattice physical quark mass squared 
divided by a number of order unity. Each small scale ratio slows down
inversion independently and the effect compounds in the worst case. 

Thus, as far as we can see, at the numerical level, there 
are no {\sl a priori} advantages to choosing true domain wall fermions
over overlap fermions in the context of QCD. In both formulations
one faces similar numerical obstacles, and the overlap,
to say the least, does not fare any worse than domain wall fermions. 
At the analytical level we are convinced
that an approach based on the overlap (or any other efficient 
replacement of the overlap Dirac operator 
that might be found in the future) is superior at presently
attainable gauge couplings in numerical QCD. Perturbation theory
is more transparent to interpret and technically less complex in the
overlap version. The chirality violating effects associated with
the number of extra fields are much more explicit and therefore
their impact should be easier to trace through.

Algorithmically we have been looking only under the lamp post and we are
far from having exhausted the options there. Nevertheless, I feel that
true domain wall fermions are an inefficient way to incorporate
the new ideas about lattice chirality into practical QCD simulations. 
For other recent work on similar issues the reader is referred to~\cite{urs-rob}.

I would like to urge you to use your imagination: there must exists much better
ways than the ones we have tried until now.

\section{ACKNOWLEDGMENTS}

This research was supported in part by DOE grant DE-FG02-96ER40949
and by a Guggenheim fellowship. I would like to thank L. Baulieu and the
entire group at LPTHE for their hospitality and support. 

I also am very grateful to the organizers for inviting me to participate
and creating a relaxed and stimulating workshop in such beautiful a location.

\end{document}